\documentclass[final,5p,times,twocolumn]{elsarticle}

\usepackage{amssymb}
 \usepackage{amsmath}
\usepackage{lineno}

\journal{Nuclear Instruments and Methods in Physics Research A }

\begin{document}

\begin{frontmatter}



\title{An accurate Rb density measurement method for a plasma wakefield accelerator experiment using a novel Rb reservoir}
\author{E. \"{O}z, F. Batsch, P. Muggli}
\address{Max Planck Institute for Physics, Munich, Germany.}
\begin{abstract}
A  method  to accurately measure the density of Rb vapor  is described. 
We plan on using this method for the Advanced Wakefield (AWAKE)~\cite{bib:awake} project at CERN
, which will be the world's first proton driven plasma wakefield experiment.  The method is similar to the hook~\cite{bib:Hook} method and has been described in great detail in the work by W. Tendell Hill et. al.~\cite{bib:densitymeter}. In this method a cosine fit is applied to the interferogram to obtain a relative accuracy on the order of $1\%$ for the vapor density-length product. A  single-mode, fiber-based, Mach-Zenhder interferometer will be built and used near the ends of the 10 meter-long AWAKE plasma source to be able to make accurate relative density measurement between these two locations. This can then be used to infer the vapor density 
gradient along the AWAKE plasma source and also change it to the value desired for the plasma wakefield experiment.  Here we describe the plan in detail and show preliminary results obtained using a prototype 8 cm long novel Rb vapor cell.
\end{abstract}
\end{frontmatter}
%
%
\section{Introduction}
\begin{figure}[htbp]
\centering\includegraphics[width=1\linewidth]{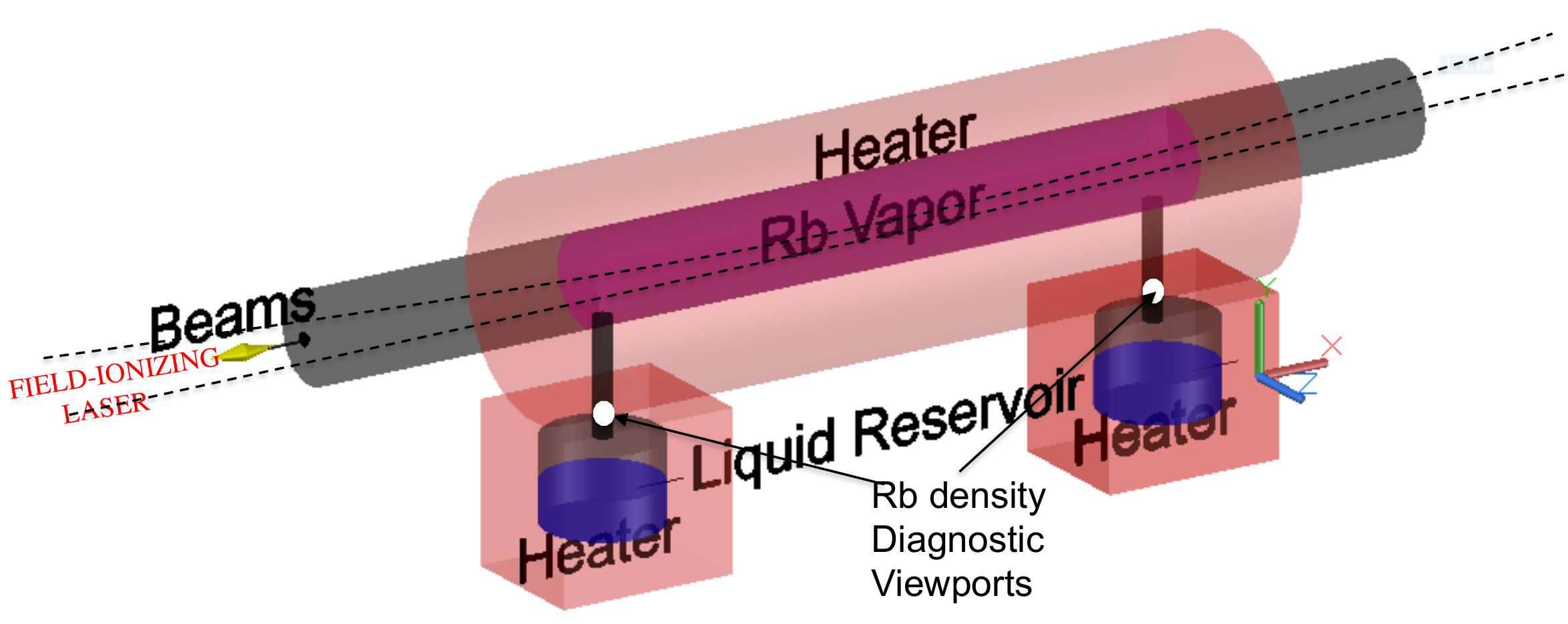}
\caption{Simple schematic of the AWAKE plasma source.}
\label{fig:awakeplasmasource} 
\end{figure}

The AWAKE collaboration will be conducting the world's first proton driven plasma wakefield experiment at CERN starting in 2016. A unique plasma source~\cite{bib:plasmasource},\cite{bib:eozipac2014} is required for this experiment. It consists of a  $\sim10$ m long rubidium (Rb) vapor confined in a 4 cm diameter tube. A simple schematic is shown in Fig. \ref{fig:awakeplasmasource}. The vapor is fully laser ionized (first Rb electron), therefore the plasma density is  equal to the vapor density. The Rb reservoirs located at both ends continuously provide Rb vapor. The middle 10 m section consists of a fluid heat exchanger and keeps the temperature (T) within $T\pm0.5$ K along the source; therefore, the vapor density profile (with a constant positive, negative or no gradient, also the plasma density profile) along the plasma source is determined by the Rb vapor flux from the ends or by the temperature uniformity in absence of flow. By measuring the Rb density accurately at the ends  one can infer the value of the density gradient.

\section{The Method}
\begin{figure}
\centering\includegraphics[width=1\linewidth]{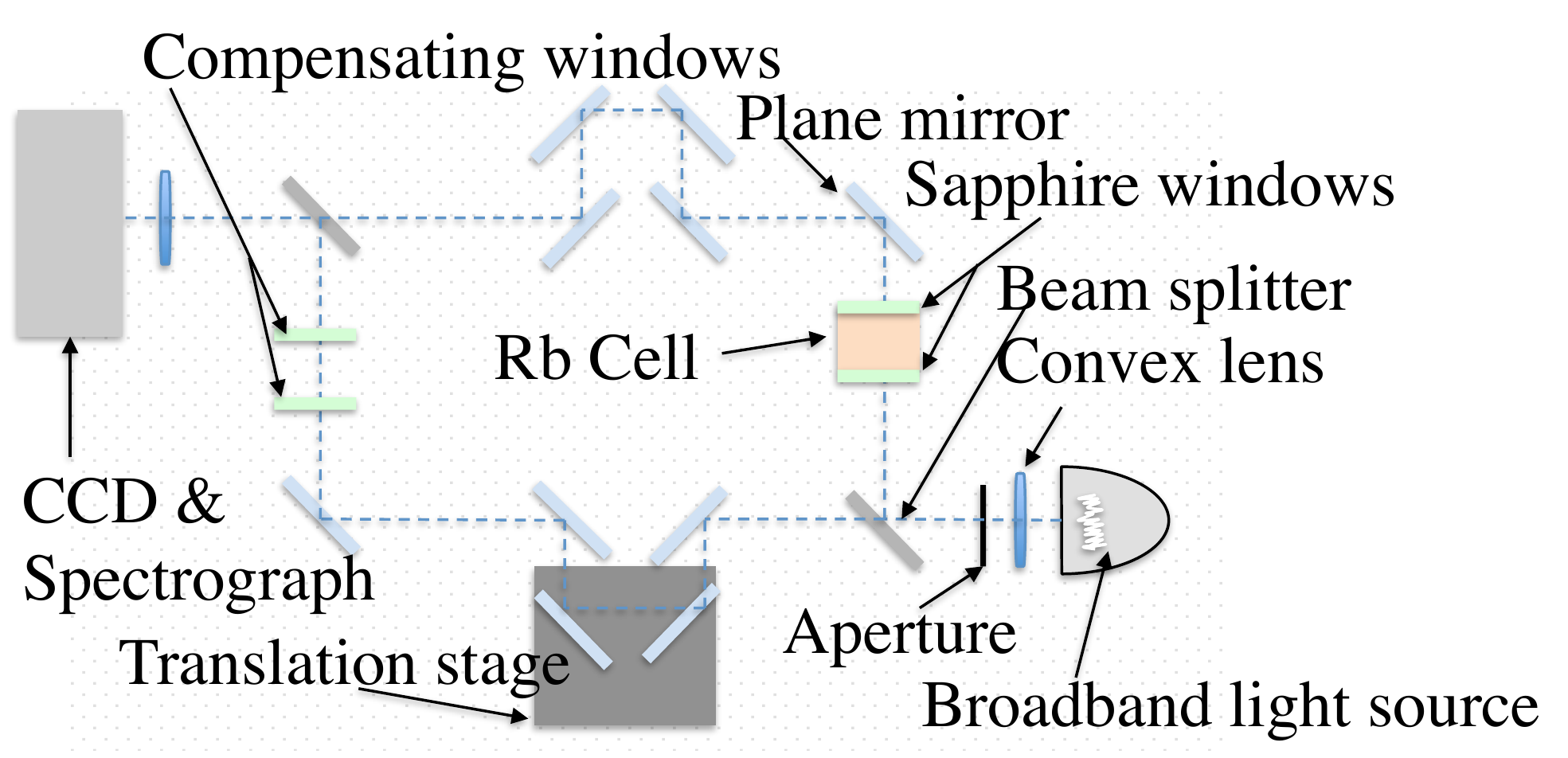}
\caption{ Schematic of the Mach-Zehnder interferometer with two delay lines.}
\label{fig:machzehnder} 
\end{figure}

\begin{figure}
\centering\includegraphics[width=1\linewidth]{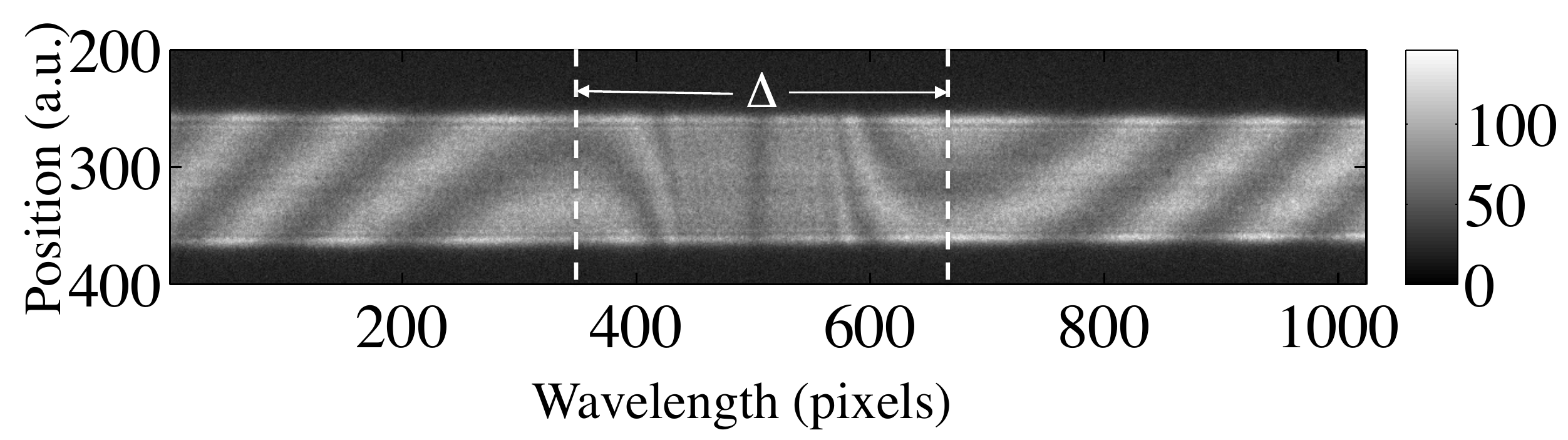}
\caption{Measured spectrum of Rb from the Mach-Zehnder interferometer set-up. Spectrograph dispersion is 0.0118 nm/pixel. The separation between the hooks is $\Delta$, from which the vapor density can be determined\cite{bib:Hook},\cite{bib:eozipac2014}.}
\label{fig:hooks} 
\end{figure}

\begin{figure}
\centering\includegraphics[width=1\linewidth]{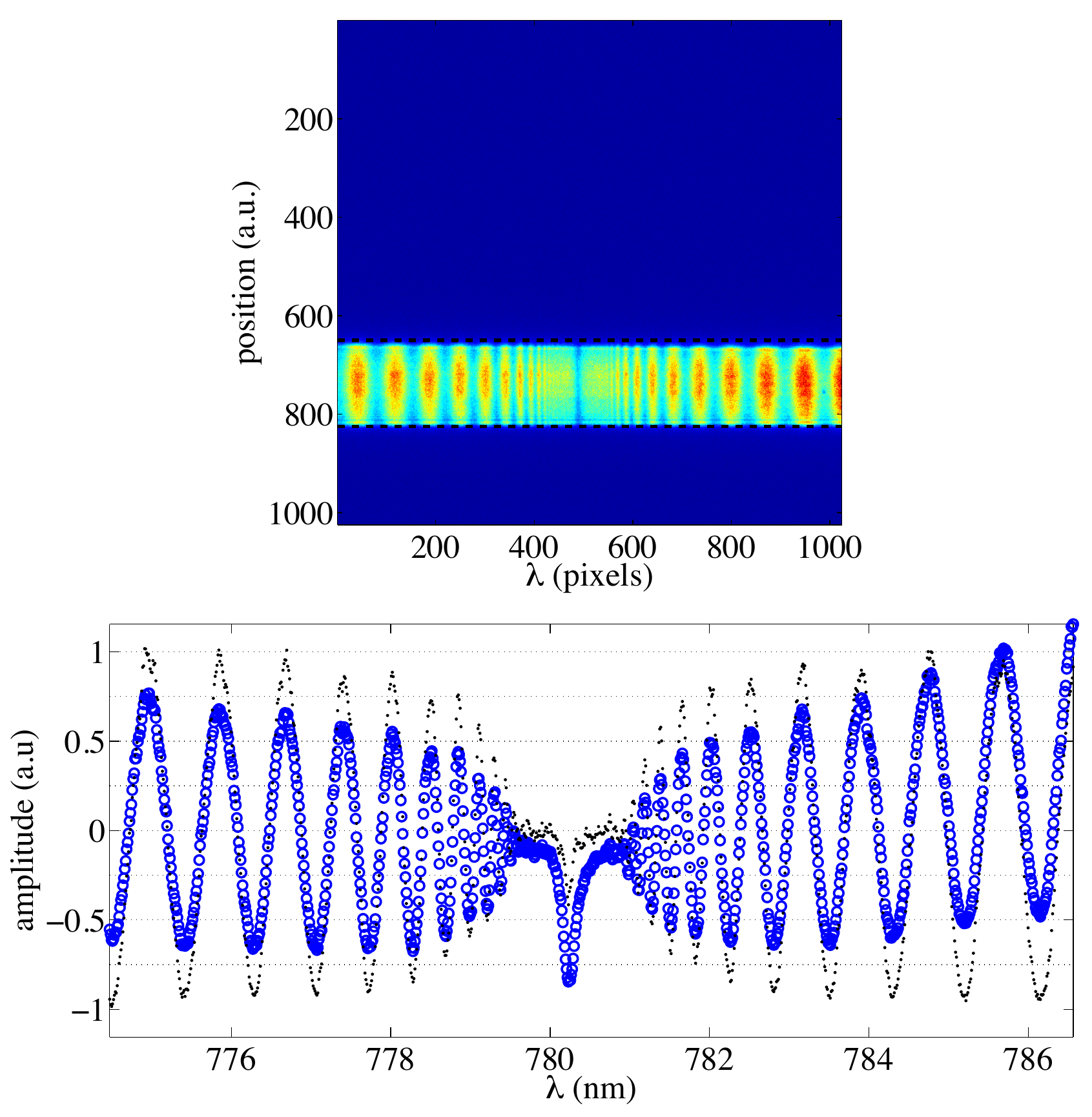}
\caption{(a) Single mode fiber Mach-Zehnder interferogram taken at a given temperature with an 8 cm long Rb vapor. (b) The vertically summed profile of the image where blue circles are the rough data and the black dots are the profile after the single arm intensities are taken into account. }
\label{fig:modulation} 
\end{figure}

Density measurements of alkali metal vapors can be made by using the anomalous dispersion around the atomic transition from their ground state and in the visible range. A Mach-Zehnder type interferometer is used (see Fig. \ref{fig:machzehnder}), where a beam of white light (i.e. broad spectrum visible light)
is split and travels through the two arms of the interferometer and is sent to an optical imaging spectrometer.  Near the resonance line, the light is subject to a large phase change that  is caused by the significant change of index of refraction of the alkali metal vapor. Therefore, when an alkali metal vapor column is placed in one of the arms of the interferometer the vapor density from the interferogram can be determined to a high degree of accuracy. If the two arms of the interferometer are not parallel then the interfering beams have different phase shifts for different positions along the spectrometer slit (vertical axis of the interferogram) and the interferogram pattern will look like that of Fig. \ref{fig:hooks}. Because of the shape of the interferogram this method is called the hook method. The wavelength difference between the hooks in this 2D interferogram is related to the vapor density. If the arms of the interferometer are parallel then one obtains vertical fringes as shown in Fig. \ref{fig:modulation}. The vapor density can be determined by making a cosine fit to the 1-d profile obtained from the horizontal line-out of the interferogram or by summing the interferogram vertically. Note that in this case an imaging spectrometer is not necessary. Since more information is used with this method to extract the density when compared to the hook method, one can reach higher accuracy. We now look at the fit function in greater detail.

\par
\begin{figure}
	\centering
		\includegraphics[width=0.5\textwidth]{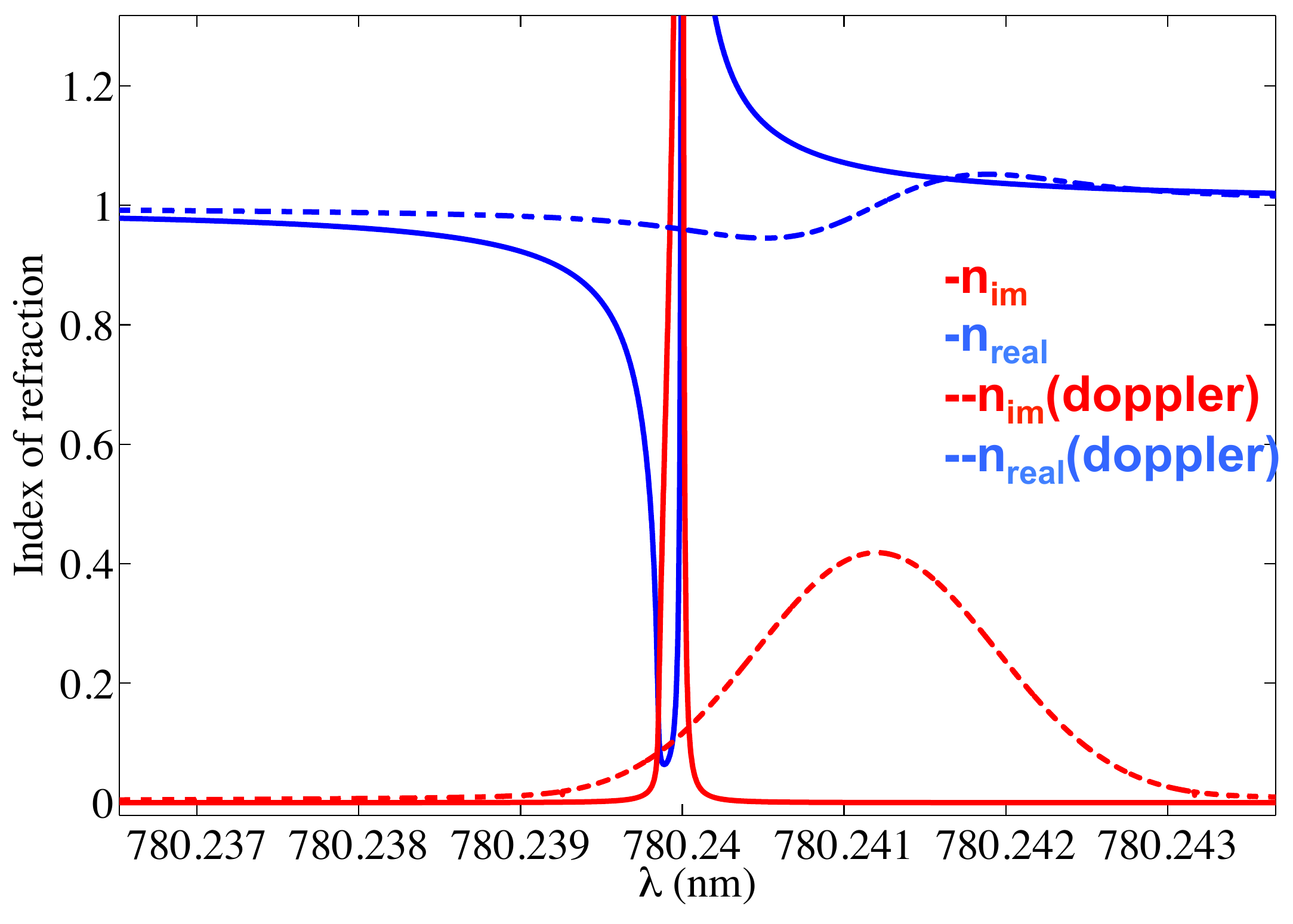}
	\caption{Real and imaginary parts of the index of refraction is plotted with (dashed lines) and without (continuous lines) Doppler broadening. The y axis is cut in order to better see the Doppler broadening effect.}
	\label{fig:doppler}
\end{figure}

The total intensity of the two interfering light beams is given by \[ I(\lambda) = I_1(\lambda)+I_2(\lambda)+2\sqrt{I_{1}(\lambda)I_{2}(\lambda)}\cos(\Delta \phi)\]
where \[\Delta \phi = \frac{2\pi}{\lambda}((n-1)l\pm\epsilon).\] Here, $\lambda$ is the wavelength, l is the length of the vapor column, $\epsilon$ is the optical path length difference between the two arms without the alkali metal vapor, the sign depends whether the vapor is in the long or short arm of the interferometer. The complex index of refraction of a gas at density $N_i$, assuming all the electrons are in the ground state (i.e. no excitation), in mks units is given by
\begin{equation}
n=\sqrt{1+\sum \frac{f_{i} e^2 N_i/(m_e \epsilon_0)}{\omega_i^2-\omega^2-i\gamma \omega}}
\label{eqn:indexmks}
\end{equation}
and in cgs
\begin{equation}
n=\sqrt{1+\sum \frac{4\pi f_{i} e^2 N_i/m_e}{\omega_i^2-\omega^2-i\gamma \omega}}
\label{eqn:indexcgs}
\end{equation}
The constants are given in table \ref{table:constants}. The sum is over all the atomic lines; however,  the main contributions come from two lines at $\lambda_1\sim780$ 
(known as the D2 line) and $\lambda_2\sim795$ nm (known as the D1 line).  Therefore we write
with $\omega_i=2\pi c/\lambda_i$
\begin{equation}
n=\sqrt{1+\frac{f_{1} e^2 N/(m_e \epsilon_0)}{(2\pi c/\lambda_1)^2-\omega^2-i\gamma_1 \omega}+
\frac{f_{2} e^2 N/(m_e \epsilon_0)}{(2\pi c/\lambda_2)^2-\omega^2-i\gamma_2 \omega}}
\label{eqn:indexrb}
\end{equation}
where $N$ is the electron number density of the ground state and is equal to the vapor density.
\begin{table}[htdp]
\caption{Various constants}
\begin{center}
  \begin{tabular}{@{} ccc @{}}
    \hline
\tiny    constant & \tiny mks & \tiny cgs  \\ \hline
\tiny    mass,$m_e$ & \tiny 9.10938215$\times 10^{-31}$ kg & \tiny 9.10938215$\times 10^{-28}$ g\\ \hline
\tiny    charge,e &\tiny 1.602176487$\times 10^{-19}$ coulombs&\tiny 4.8032068$\times 10^{-10}$ esu\\ \hline
\tiny    permittivity, $\epsilon_0$ &\tiny 8.8542$\times 10^{-12}$ Farad $m^{-1}$& \\ \hline
\tiny    $\lambda_1$~\cite{bib:rbwave}&\tiny780.24$\times 10^{-9}$ m &\tiny780.24$\times 10^{-7}$ cm \\ \hline
\tiny     $\lambda_2$~\cite{bib:rbwave} &\tiny794.98$\times 10^{-9}$ m& \tiny794.98$\times 10^{-7}$ cm \\ \hline
\tiny    oscillator strength, $f_{1}$~\cite{bib:rbosc}&\tiny0.69577& \tiny0.69577 \\ \hline
\tiny     oscillator strength, $f_{2}$~\cite{bib:rbosc}&\tiny0.34231 &\tiny 0.34231 \\ \hline
\tiny     level life time, $\tau_1$~\cite{bib:lifetime} & \tiny26.24$\times 10^{-9}$ s & \tiny26.24$\times 10^{-9}$ s  \\ \hline
\tiny     level life time, $\tau_2$~\cite{bib:lifetime}  & \tiny27.70$\times 10^{-9}$ s & \tiny27.70$\times 10^{-9}$ s  \\ \hline
\tiny    atomic absorption coefficient, $\gamma_1=1/\tau_1$ &\tiny 3.811$\times 10^{7}$&\tiny3.811$\times 10^{7}$ \\ \hline
 \tiny   atomic absorption coefficient, $\gamma_2=1/\tau_2$ & \tiny3.61$\times 10^{7} $ $s^{-1}$&\tiny3.61$\times 10^{7}$ $s^{-1}$\\ \hline
    \hline
  \end{tabular}
\end{center}
\label{table:constants}
    \end{table}
The formula above ignores line broadening. For our temperature range ( $<\sim$ 200 $^{\circ}$C ) the Doppler broadening effect dominates over other broadening mechanisms. The index of refraction including Doppler broadening is ~\cite{bib:doppler}:
\begin{equation}
\begin{split}
n=\{1+  \int_{-\infty}^{+\infty}  \frac{1}{2\sqrt{\pi}\Omega_1} \frac{e^{-\frac{\Omega^2}{\Omega_{1}^2}} d \Omega f_{1} e^2 N/(m_e \epsilon_0)}{(\omega_1+\Omega)^2-\omega^2-i\gamma_1 \omega}+ \\
\int_{-\infty}^{+\infty} \frac{e^{-\frac{\Omega^2}{\Omega_{2}^2}} d \Omega f_{2} e^2 N/(m_e \epsilon_0)}{(\omega_2+\Omega)^2-\omega^2-i\gamma_2 \omega}\frac{1}{2\sqrt{\pi}\Omega_2}\}^{1/2}
\end{split}
\label{eqn:doppler}
\end{equation}
where $\Omega_1 = \frac{2\pi c}{\lambda_1} \sqrt{\frac{5kT}{3M}} $, T is the temperature of Rb, M is the atomic mass of Rb, $M=85.4678/6.02214\times10^{23} $ g, k is the Boltzman constant, $k=1.3807 \times 10^{-23}$  $JK^{-1}$. For $T=500$ K, $\Omega_1=2.30\times 10^{9}$ $s^{-1}$ and $\Omega_2=2.26\times 10^{9}$ $s^{-1}$,  c is the speed of light, $c=299792458$ m/s~\cite{bib:CRC}. 
The doppler coefficients are determined by \[\int_{-\infty}^{+\infty}  \frac{1}{2\sqrt{\pi}\Omega_1} e^{-\frac{-\Omega^2}{\Omega_{1}^2}} d \Omega +\int_{-\infty}^{+\infty}  \frac{1}{2\sqrt{\pi}\Omega_2} e^{-\frac{-\Omega^2}{\Omega_{2}^2}} d \Omega=1.\] 
There is no analytical solution to the integral therefore we numerically integrate it after separating the complex and real parts. We plot the doppler broadened index and susceptibility in the range of $\lambda_1$ in Fig. \ref{fig:doppler}. 

As can be seen from the Fig. \ref{fig:doppler} the effect of the doppler broadening on the real part of index of refraction more than 1 nm away from the absorption line on both sides is negligible. The absorption also plays no role for the purpose of this calculation so far away from the resonance line; therefore, the index of refraction in cgs is approximately given by
 \[n(\lambda)-1=\frac{e^{2}Nf_{1}\lambda_{1}^3}{4\pi m_{e}(\lambda-\lambda_{1})} +\frac{e^{2}Nf_{2}\lambda_{2}^3}{4\pi m_{e}(\lambda-\lambda_{2})}.\] 

In order to determine the density-length product one needs to record three spectra, i.e. $I_1(\lambda)$, $I_2(\lambda)$, $I_T(\lambda)$ and then apply a cosine fit to S($\lambda$) given below. The amplitude, A,  is also kept as a fitting parameter because of the imperfect of the fringes.  
\begin{eqnarray}
S(\lambda)=A \frac{I-I_1-I_2}{2\sqrt{I_1 I_2}} \\
S(\lambda)=A\cos(\frac{2\pi}{\lambda}(\frac{e^{2}Nf_{1}l\lambda_{1}^3}{4\pi m_{e}(\lambda-\lambda_{1})}+  \frac{e^{2}Nf_{2}\lambda_{2}^3}{4\pi m_{e}(\lambda-\lambda_{2})}  \pm \epsilon))
\label{eqn:fitfunction}
\end{eqnarray}

\section{Experimental Set-up and Results}

\begin{figure}
\centering\includegraphics[width=1\linewidth]{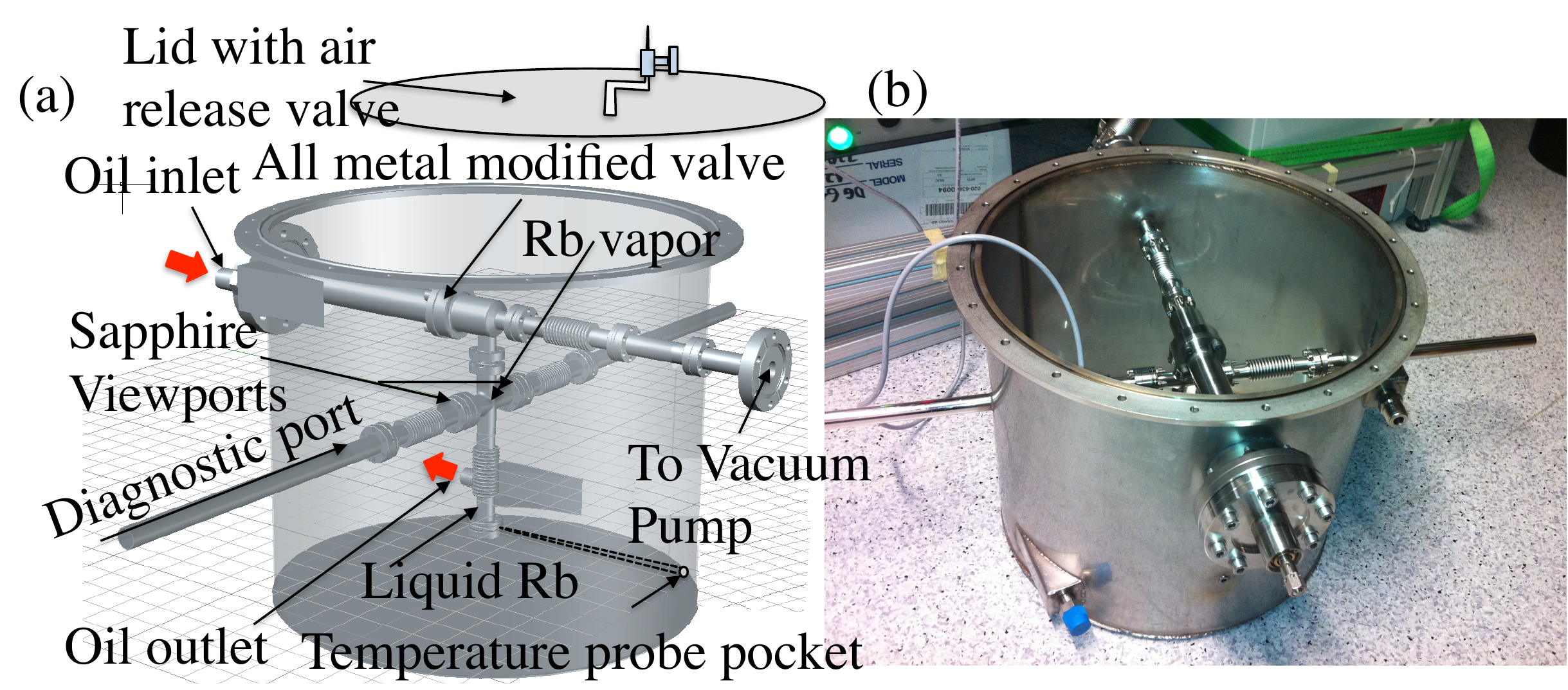}
\caption{(a)Drawing of the oil heated Rb reservoir, oil bath with heater not shown.(b) Photo of the Rb reservoir. }
\label{fig:rbreservoir} 
\end{figure}

\begin{figure}
\centering\includegraphics[width=1\linewidth]{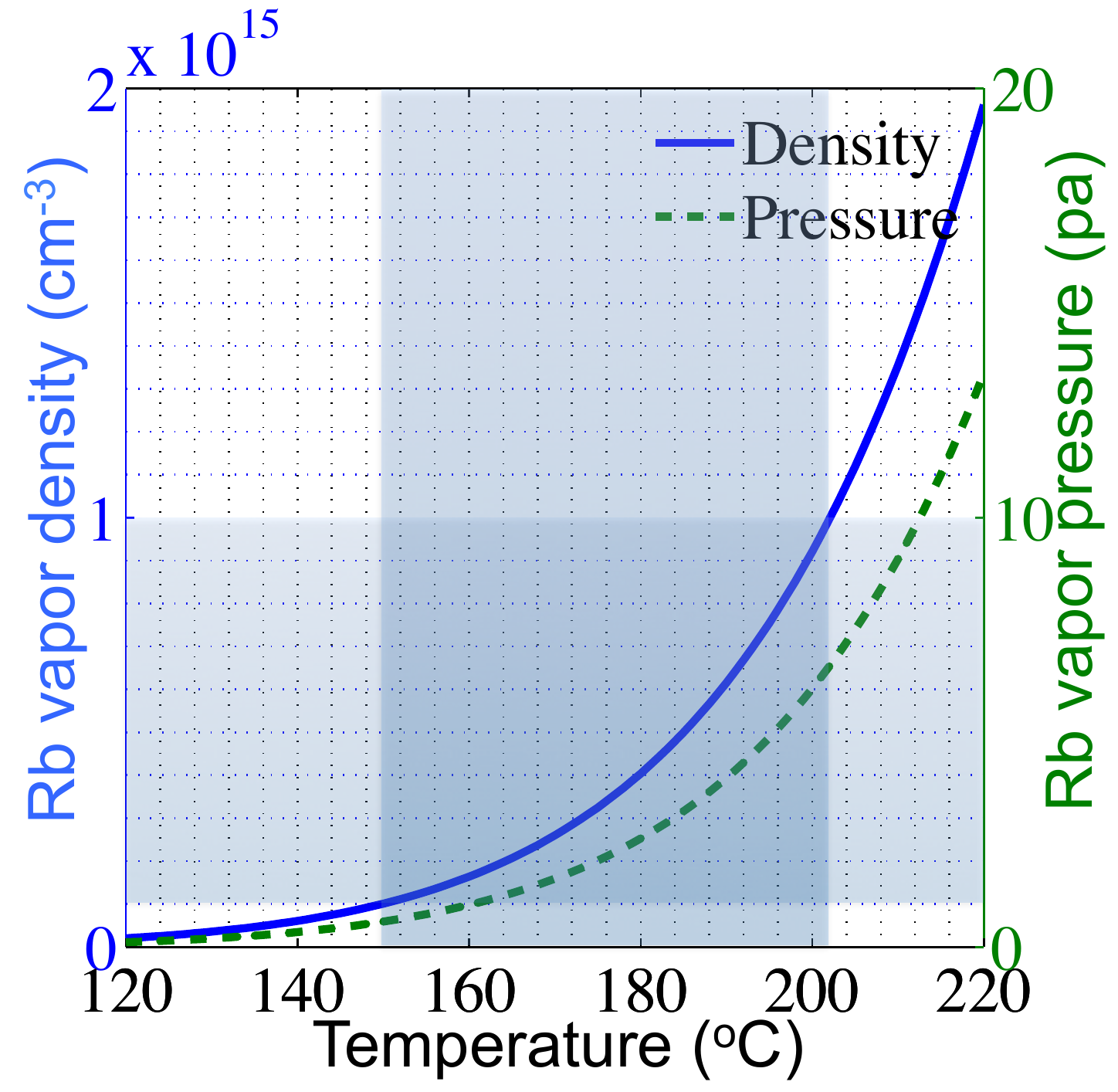} 
\caption{Rubidium vapor density (blue line) and vapor pressure (green dashed line) as a function of temperature. Region between $1\times10^{14}$~cm$^{-3}$ and $1\times10^{15}$~cm$^{-3}$, and the corresponding temperature shows the parameter range of interest for the PDPWFA.}
\label{fig:VaporPressure}
\end{figure}

\par
 A novel oil heated reservoir was developed and used for this study (see Fig. \ref{fig:rbreservoir}).  An oil bath with a circulating pump is used to heat the Rb reservoir. Being immersed in circulating oil keeps the temperature of the entire system uniform and one can determine the temperature of the Rb liquid, which determines the vapor density, using the probe inserted in the temperature pocket (a thin stainless steel tube) as shown in the schematic (Fig. \ref{fig:rbreservoir} (a)). The reservoir contains an in-house modified all metal angle valve. The modification allows additional oil sealing and the elongated arm reduces heat losses.  The reservoir is sealed using the lid, filled with silicon oil and then insulated against heat losses using insulation wool on the outside. When heated the Rb vapor fills the region of the DN 16 cross which has sapphire diagnostic viewports.  
\par
\begin{figure}
\centering\includegraphics[width=1\linewidth]{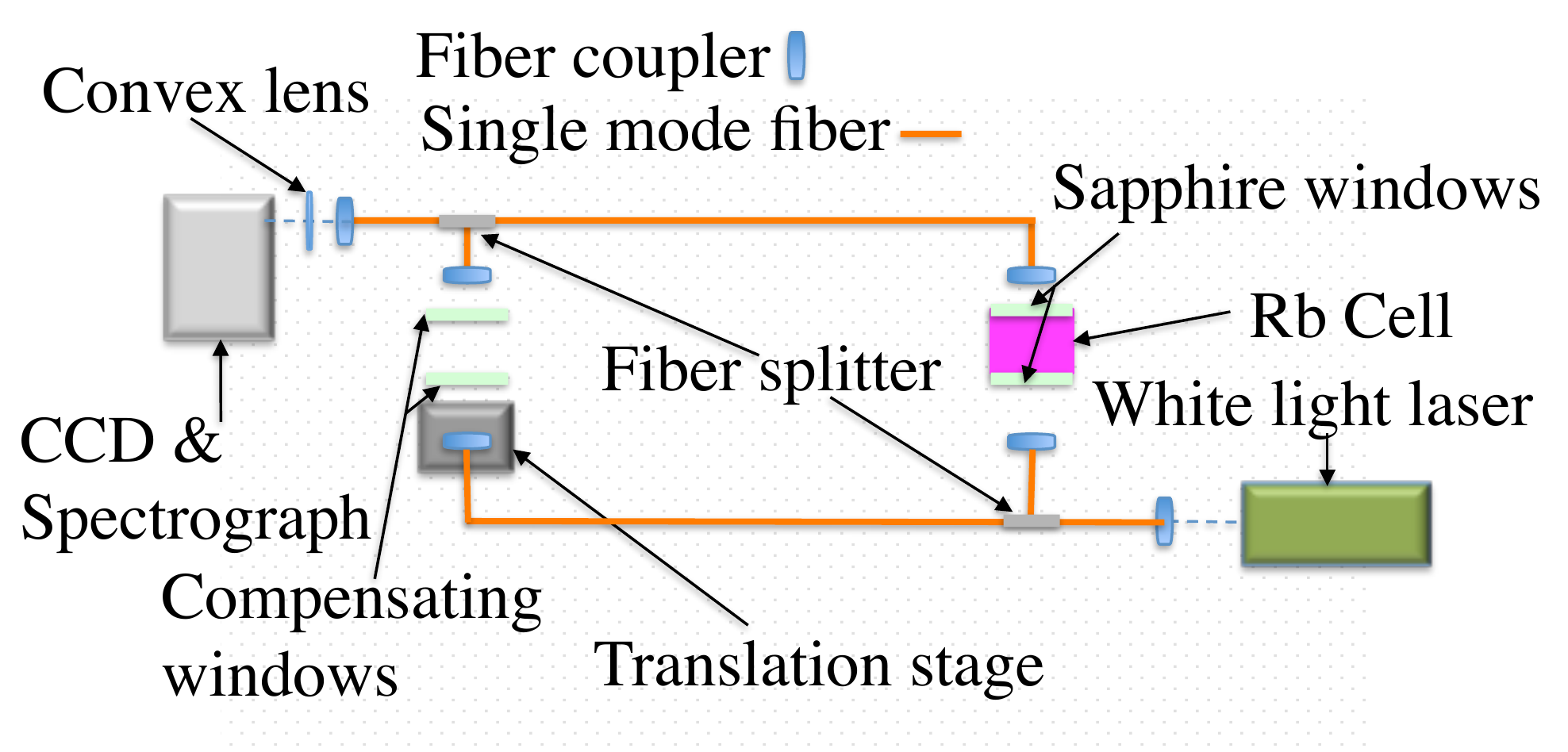}
\caption{Single mode fiber based Mach-Zehnder interferometer with a white light laser. Rb vapor cell is placed in one of the arms and the interference pattern is recorded using a spectrograph with a CCD. The translation stage allows for the adjustment of relative path difference between the two arms of the interferometer.}
\label{fig:fibermachzehnder} 
\end{figure}

For the measurements, the reservoir is placed in one of the arms of the single mode fiber-based, Mach-Zehnder set-up as shown in Fig. \ref{fig:fibermachzehnder}. A white light laser  is used as the light source. The high intensity laser is chosen since for the AWAKE experimental setup light is transported away from the radiation area using single mode fibers. The fibers used for this wavelength range ($\lambda$ $\sim$ 780 nm) have a core diameter of 3.5 $\mu$m. Using a multimode fiber causes speckles (random intensity fluctuations) in the interferogram making it unusable~\cite{bib:fabianthesis}.

\begin{figure}
\centering\includegraphics[width=1\linewidth]{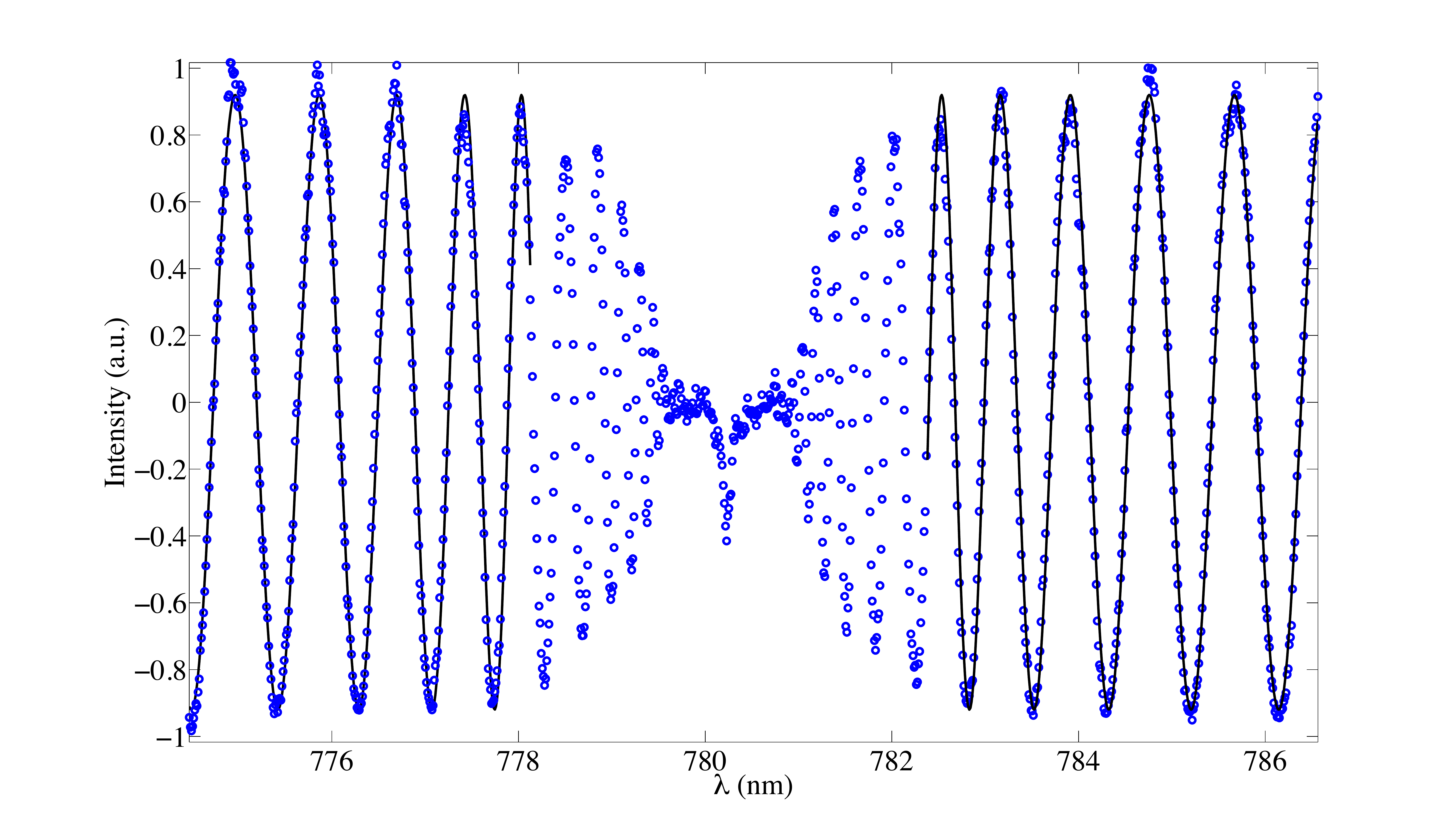}
\caption{Least square fit (solid black line) to the profile (blue circles).  Same profile is shown in Fig. \ref{fig:modulation} (b) as a black dotted line.}
\label{fig:fit} 
\end{figure}

\par
Three spectra were recorded at T$=$ 188.9 $^{\circ}$C. The spectra corresponding to I$_T$ and the vertically summed profile are shown in Fig. \ref{fig:modulation}. Note that near the resonance when the period is comparable to the resolution of the spectrograph ($\sim$ 0.04 nm) the amplitude starts to drop (Fig. \ref{fig:modulation}). This is due to the convolution of the instrument function with the signal (the cosine function).  We obtain the function S in eq. \ref{eqn:fitfunction} using the three spectra and a nonlinear least square fit is applied to the the region sufficiently away from the absorption line (see Fig. \ref{fig:fit}).  One can also extend the fit to include the central region by using a convolved fit function. The density corresponding to the vapor pressure curve for this temperature is 5.9 $\times$ $10^{14}$ cm$^{-3}$ and the fit gives 6.0 $\times$ $10^{14}$ cm$^{-3}$, which is within expected absolute accuracy (5$\%$) of vapor curve expression~\cite{bib:CRC}. For the AWAKE experiment, absolute accuracy is not so important since the required quantity for determining the density gradient is a relative measurement between the two locations. 
\par
We assess the accuracy of the density measurements in two ways. One is  from a single spectra including the experimental errors in y axis as follows. We generate multiple random data sets (each corresponds to a virtual experiment) according to the experimental errors. Uncertainty in the y axis is the poisson noise which is the square root of the number of photoelectrons (note that number of CCD counts must be converted to photoelectron count, the specified number by the manufacturer is $\sim$ 1 counts/photoelectron).   Then non-linear fitting is applied to each random data set,  the standard deviation ($\sigma$ $\sim$0.2$\%$) that results from these data sets gives the uncertainty in the density measurement. This statistical analysis of calculating uncertainty using experimental errors from a single spectra is ideally equivalent to taking multiple shots and  taking the standard deviation of the data set. Since the multiple shots will be randomly distributed according to the experimental uncertainties.  Doing so for multiple  spectra taken consecutively  we find that the density fluctuation remains small (namely $\sigma$ $\sim$0.5$\%$ ). Vibrations contribute random phase shifts, but this is handled by the fitting algorithm. 

\section{Implementation to the AWAKE Proton Driven Plasma Wakefield Accelerator Experiment}
For the AWAKE experiment two ports at the ends of the plasma source are planned to be used. Each Rb vapor source has an independent temperature controller. The relative accuracy of the temperature measurement between the two reservoirs is planned to be around $\pm$0.05 K (using platinum resistor temperature sensors, pt-111, or silicon diode sensors, DT670). Although the temperature can be adjusted to such high accuracy, the temperature of the coldest point of liquid Rb may not correspond to the set temperature. Since the coldest point determines the vapor density this introduces a challenge. This can be  handled by combining temperature measurement with the high accuracy interferometric density measurement. To see this we calculate the fractional change in the density ( $\frac{dn}{n}$) corresponding to the fractional temperature adjustment $\frac{dT}{T}$ using the Clasius-Clapeyron relation. The vapor pressure is given by 
\[P = P_i e^{\frac{h}{k_ B}(\frac{1}{T_i}-\frac{1}{T})} \] and 
\begin{equation}
 \frac{dP}{P} = \frac{h}{k_B}\frac{dT}{T}
\label{eqn:cp}
\end{equation}
where $ h=1.1\times 10^{-19} $ joules is the heat of vaporization per Rb atom and T is the temperature of the liquid. According to the ideal gas equation: \[ p=nkT \] where p  is pressure and n is density. Therefore at constant temperature \[\frac{dp}{p}=\frac{dn}{n}\] and using eq. \ref{eqn:cp} one can calculate the effect of change of liquid Rb temperature on the density.  For example for the optimum AWAKE plasma density $n=7\times10^{14}$ cm$^{-3}$ which corresponds to a temperature T$\sim 466$ K (given by the vapor pressure curve of Rb~\cite{bib:CRC})  (see Fig. \ref{fig:VaporPressure}), we get: \[\frac{dn}{n} =17.09\frac{dT}{T}.\] Therefore, a temperature change of 0.1 K, changes the Rb vapor density by 0.36$\%$.  Hence, using this temperature control knob which is capable of such fine adjustments with the interferometric measurement described here one can adjust the two Rb sources relative two each other and independently verify that the relative density measurement is consistent.

\par
Note that in this method one measures the density length product. Therefore the length of the vapor column of the two different reservoirs needs to be known to a high accuracy.
This can be done by using commercial mechanical micrometers with 1 micrometer accuracy. Since the vapor column is planned to be a few centimeters long, the uncertainty in length measurement is negligible.

\section{Conclusion}

In conclusion, an interferometric method to measure Rb vapor density length product to high relative accuracy (n $\pm$ 0.5$\%$) is presented. Initial measurements performed using a novel Rb reservoir confirms the feasibility of the method. This diagnostic is planned to be used for the world's first proton driven plasma wakefield experiment, AWAKE. The unique 10 m long AWAKE plasma source density profile can be determined using this method. 
 
\section{Acknowledgements}
We acknowledge many fruitful discussion with Prof. Dr. Allen C. Caldwell, the members of the AWAKE collaboration  and thank Mr. Thomas Haubold and Gennadiy Finenko from Max Planck Institute for Physics for their expert technical help.
\bibliographystyle{elsarticle-num}
\bibliography{rbdensity}

\begin{thebibliography}{10}
\expandafter\ifx\csname url\endcsname\relax
  \def\url#1{\texttt{#1}}\fi
\expandafter\ifx\csname urlprefix\endcsname\relax\def\urlprefix{URL }\fi
\expandafter\ifx\csname href\endcsname\relax
  \def\href#1#2{#2} \def\path#1{#1}\fi

\bibitem{bib:awake}
R.~Assmann, et.~al. (AWAKE~Collaboration), Proton-driven plasma wakefield
  acceleration: a path to the future of high-energy particle physics, Plasma
  Physics and Controlled Fusion 56~(8) (2014) 084013.

\bibitem{bib:Hook}
W.~C. Marlow, Hakenmethode, Appl. Opt. 6~(10) (1967) 1715--1724.
\newblock \href {http://dx.doi.org/10.1364/AO.6.001715}
  {\path{doi:10.1364/AO.6.001715}}.

\bibitem{bib:densitymeter}
W.~T. Hill, Column-density meter: a high precision technique for measuring
  line-of-sight vapor densities, Appl. Opt. 25~(23) (1986) 4476--4482.

\bibitem{bib:plasmasource}
E.~\"{O}z, P.~Muggli, A novel rb vapor plasma source for plasma wakefield
  accelerators, Nuclear Instruments and Methods in Physics Research Section A:
  Accelerators, Spectrometers, Detectors and Associated Equipment 740 (2014)
  197--202.

\bibitem{bib:eozipac2014}
E.~\"{O}z, F.~Batsch, P.~Muggli, A novel laser ionized \textsc{R}b plasma
  source for plasma wakefield accelerators, Proc. IPAC2014 104 (2014)
  1522--1524.

\bibitem{bib:rbwave}
G.~Barwood, P.~Gill, W.~Rowley, Frequency measurements on optically narrowed
  rb-stabilized laser diodes at 780 nm and 795 nm, Applied Physics B 53~(3)
  (1991) 142--147.
\newblock \href {http://dx.doi.org/10.1007/BF00330229}
  {\path{doi:10.1007/BF00330229}}.

\bibitem{bib:rbosc}
D. \textsc{S}teck, http://steck.us/alkalidata/.

\bibitem{bib:lifetime}
U.~Volz, H.~Schmoranzer, Precision lifetime measurements on alkali atoms and on
  helium by beam-gas-laser spectroscopy, Physica Scripta 1996~(T65) (1996) 48.

\bibitem{bib:doppler}
K.~G. Libbrecht, M.~W. Libbrecht, Interferometric measurement of the resonant
  absorption and refractive index in rubidium gas, American Journal of Physics
  74~(12) (2006) 1055--1060.

\bibitem{bib:CRC}
CRC Handbook of Chemistry and Physics, 87th Edition (CRC Handbook of Chemistry
  \& Physics), 2014.

\bibitem{bib:fabianthesis}
F.~Batsch, Measurement of alkali metal vapor density using interferometric
  methods, 2014, bachelor thesis, Technical University, Munich.

\end{thebibliography}






\end{document}